\documentclass{article}                       
\usepackage{isolatin1,bezier,epsfig}
\begin{document}                                    
\newcommand{\be}{\begin{equation}}
\newcommand{\ee}{\end{equation}}
\newcommand{\bq}{\begin{quote}}
\newcommand{\eq}{\end{quote}}
\newcommand{\bi}{\begin{itemize}}
\newcommand{\ei}{\end{itemize}}
\newcommand{\ben}{\begin{enumerate}}
\newcommand{\een}{\end{enumerate}}
\newcommand{\ba}{\begin{array}}
\newcommand{\ea}{\end{array}}
\newcommand{\bea}{\begin{eqnarray*}}
\newcommand{\eea}{\end{eqnarray*}}
\newcommand{\baa}{\begin{eqnarray}}
\newcommand{\eaa}{\end{eqnarray}}

\begin{center}
{\bf\large Contribution to the ``Klaus Ruedenberg-Volume''}

\vspace*{3cm}
{\bf \Large 
Molecular calculations with {\it B} functions
}
\end{center}
\vspace{1cm}
\large
\begin{center}
E.O. Steinborn$^a$, H.H.H. Homeier$^a$, I. Ema$^b$, R. Lopez$^b$ and G. Ramírez$^b$
\end{center}
\begin{center}
$^a${\it
Institut f\"ur Physikalische und Theoretische Chemie,
Universit\"at Regensburg, \\
D-93040 Regensburg, Germany.}

$^b${\it
Departamento de Química Física Aplicada, Facultad de Ciencias C-XIV,
Universidad Autónoma de Madrid, 28049 Madrid, Spain.}
\end{center}

{\bf ABSTRACT}

A program for molecular calculations with $B$ functions is
reported and its performance is analyzed. 
All the one- and two-center integrals, and
the three-center nuclear attraction integrals are computed by direct
procedures, using previously developed algorithms.  The three-
and four-center electron repulsion integrals are computed by means of
Gaussian expansions of the $B$ functions. A new procedure for obtaining
these expansions is also reported.
Some results on full molecular calculations are included to show the
capabilities of the program and the quality of the $B$ functions to
represent the electronic functions in molecules.

\section{Introduction}

Although there is a general agreement in the best quality of
exponential type orbitals (ETO) for describing the electronic
function in atoms and molecules \cite{preced},
the difficulties in solving some
molecular integrals (in particular the three- and four-center
two-electron repulsion integrals) have strongly restricted their
use in molecular calculations, in benefit of the more easy to
handle Gaussian type orbitals (GTO).

In a previous article \cite{calcSTO}, a program for
molecular calculations with Slater type orbitals (STO)
has been reported. In that program,
some of the molecular integrals (all the one- and two-center and the
three-center nuclear attraction) were computed by direct
algorithms, and the remaining ones by means of Gaussian
expansions of the STO. The aim of that program was
to provide a bridge and a reference for further works in this field,
mainly to those dealing with the direct calculation of all the
integrals.
The present work has been developed in the same spirit, using in this
case a different type of ETO, the reduced Bessel functions 
($B$ functions) that have been already proposed by Steinborn et
al. \cite{steinB1,steinB2} for molecular calculations.

In the real case, the $B$ functions are
defined as: 
\be
B_{n,l}^m(\alpha,\mathbf{r}) = [2^{n+l} \; (n+l)!]^{-1}
\hat{k}_{n-1/2}(\alpha r) \; (\alpha r)^l \; z_l^m(\theta,\phi)
\ee
where
\be
\hat{k}_{\gamma}(z) = \sqrt{\frac{2}{\pi}} \; z^\gamma \;
K_{\gamma}(z)
\ee
are the corresponding reduced Bessel functions, with $K_\gamma$ being
the Macdonald function \cite{BesselK},
$z_l^m$ are real spherical harmonics:
\be \label{eq:1.1}
z_l^m(\theta,\phi) = 
\sqrt{\frac{(2l+1) \; (l-|m|)!}{(1+\delta_{m,0}) \; \pi \; (l+|m|)!}}
(-1)^m P_l^{|m|} (\cos \theta) \; \Phi_m(\phi)
\ee
$P_l^{|m|}$ being the corresponding Legendre
functions  \cite{Legendre}, and the functions $\Phi$ are defined as:
\be \label{eq:1.2}
\Phi_m(\phi) = \left\{
\begin{array}{ll}
\cos m \phi & \mbox{for } m \ge 0 \\
\sin |m| \phi & \mbox{for } m < 0
\end{array}
\right.
\ee
The simplicity of the Fourier transform of the $B$ functions has been
proposed as an advantage to be exploited in the development of
algorithms for molecular integrals calculations with this kind of
ETO \cite{steinB1,steinB2}.

In the next sections, the algorithms used in the program will be
summarized and some results will be reported to show the quality
of the $B$ functions for molecular calculations and to compare with
both the commonly used Gaussian functions and with the STO.

\section{Algorithms}

The present program for molecular integrals with $B$ functions closely
follows the scheme of that previously reported for the STO \cite{calcSTO}. The one- and
two-center and the three-center nuclear attraction integrals are
computed by direct procedures previously 
developed \cite{steinB1,steinB2,1e2c,2e2c,Willy,STO1e3c} and the
remaining ones are computed by means of Gaussian expansions of the $B$
functions.

The integrals computed with direct procedures are obtained
with an average cost about 0.1 milliseconds
per integral actually computed measured on a Digital AlphaServer 8400,
for an accuracy of at least twelve decimal places.

The remaining integrals have been computed by using Gaussian
expansions, as mentioned above. The expansions have been obtained by a
procedure different than the usual least-squares method. This procedure,
that is summarized in the appendix, is much easier to apply than the
least-squares, and yields expansions that tend to better approximate the
tails of the functions. 

The integrals between the Gaussian primitives
are evaluated with the algorithm 
proposed by Saunders \cite{Saunders,CPC3} slightly modified, and
a test for avoiding calculation of negligible Gaussian
contributions to a given set of integrals is also included.

The program for the calculation of the integrals has been implemented in
a modular way, so that the subroutines corresponding to each type of
integrals can be easily replaced by others when required.
The main structure can therefore be kept unaltered when trying new
algorithms for a given type of integrals. 

The package includes routines for the direct minimization of
the energy in RHF and ORHF calculations \cite{mindir} but, since the
integrals are stored in external files in a very easy-to-handle way,
the interface with other standard programs for minimizing the energy
(both in Hartree-Fock and post-Hartree-Fock calculations) is
straightforward.

\section{Results}

We have first analyzed the quality of the Gaussian expansions for
representing the $B$ functions. In Table 1 we collect the least-squares
error of the expansions for different principal quantum numbers in the
functions. We want to recall here that {\it the least-squares error
has not been used as criterion for attaining the present expansions}.

As it can be seen in the table, the quality of the expansions 
improves as the $n$ quantum number increases. This is so because the
Gaussian expansions obtained with the current procedure tend to better
reproduce the tail of the functions, but present some problems to
reproduce the peak in the origin for the functions with $n = 1$. This
causes a loss of accuracy in the integrals involving these latter
functions. In consequence, the program uses 
expansions optimized by minimizing the least-squares error 
for $\hat{k}_{1/2}$ (which
are the same as those of the STO 1s \cite{stng1s2s}), and
expansions obtained with the current procedure for the rest.

Then, we have carried out full RHF calculations on several systems to
test the performance of the program and to analyze the qualities of the
$B$ functions for molecular calculations.

The corresponding exponents of the single-zeta and
double-zeta basis sets used in the calculations are collected
in Tables 2 and 3.
These basis sets are optimized for atoms and have been taken from
references \cite{dotterweich,bases}. The geometries are summarized in
Table 4.

In Tables 5 and 6, the energy values and the computational cost are
analyzed for expansions of different lenght with both types of basis
sets. The computational cost of
the integrals calculated by direct procedures is included in the second
column of each table, and columns 3 to 8 collect both the value of the
electronic energy and the time required for the integrals computed with
the Gaussian expansions for the B-10G, B-15G and B-20G expansions.
As it can be seen in these tables, 
the direct procedures enable us to obtain highly accurate integrals in
a very fast way. This should spur for a further search on direct
procedures for all the integrals. 

Nevertheless, an accuracy sufficient for
testing purposes can be attained with a moderate cost by using not very
long expansions (such as the B-10G) of the $B$ functions. In fact, the
results obtained with these expansions are also sufficiently accurate
to give an idea about the quality of the $B$ functions in molecular
calculations. In Table 7 a comparison between the results obtained with
$B$ and STO basis sets of same length is made. The table clearly
illustrates that, despite that the energy values with STO are slightly
lower in the single zeta basis sets, the double zeta of both types
yield results of similar quality, thus confirming the capabilities of
the $B$ functions for high-quality molecular calculations.

\section{Conclusions}

A program for computing molecular integrals with $B$ functions has
been implemented and tested. The program combines techniques of direct
computation for the one- and two-center integrals and the three-center
nuclear attraction integrals with others based on Gaussian expansions for
the remaining ones. Direct procedures yield highly accurate results at
low computational cost. The Gaussian expansions lead to a lower ratio
accuracy/cost, but still acceptable for calculations in
medium size systems. The expansions obtained by the moments procedure
reported herein are good enough for all cases except the $\hat{k}_{1/2}$
function because of the peak in the origin in this latter. An expansion
based in the least-squares procedure is preferable in this case, and
has been used in the program. Finally, basis sets of $B$ functions
have a good quality for reproducing the electronic wave function in
molecules, comparable with that of STO.


\section{Appendix: Gaussian expansions of $B$ functions}

The procedure used to obtain Gaussian expansions of the B functions is
based on the fitting of the value of the functions at the origin and
the {\it moments of the unnormalized spherical part of the
B-function}, i.e., an expansion:
\be \label{eq:A.1}
\hat{k}_{n-1/2}(r) \approx \sum_{i=1}^N  c_i \; e^{-
\alpha_i \, r^2}
\ee
is chosen with the requirement that a given number of the moments of the
reduced Bessel functions are reproduced:
\be \label{eq:A.2}
\int_0^\infty dr \; r^{k-1} \; \hat{k}_{n-1/2}(r) = \sum_i c_i
\int_0^\infty dr \; r^{k-1} \; e^{-\alpha_i \, r^2}  \;\;\;\; k = 1, 2,
... 
\ee
Replacing the integrals by their final
values \cite{Grd6561.16,Grd3461.3},
the following equalities are obtained:
\be \label{eq:A.3}
2^{k+n-3/2} \; \Gamma \left( \frac{k}{2}+n \right) \;
\Gamma \left( \frac{k+1}{2} \right) 
= \frac{1}{2} \; \sum_i c_i \;
\frac{\Gamma \left( \frac{k+1}{2} \right) }{\alpha_i^{\frac{k+1}{2}}}
\ee
Introducing the definitions:
\be \label{eq:A.4aa}
Q_0 = \hat{k}_{n-1/2}(0)
\ee
\be \label{eq:A.4a}
Q_k \equiv 2^{k+n-1/2} \; \Gamma \left( \frac{k}{2}+n \right) 
\ee
and
\be \label{eq:A.4b}
\ba{lr}
r_i \equiv \frac{1}{\sqrt{\alpha_i}}
\ea
\ee
Eq.\ (\ref{eq:A.3}) can be rewritten as:
\be \label{eq:A.5}
Q_k = \sum_{i=1}^N c_i r_i^k \;\;\;\; k = 0, 1,
2N-1
\ee
Since an expansion of length $N$ implies $2N$ unknowns, $c_i$, $r_i$, a
total of $2N$ moments ($k = 0, 1, ...2N-1$) can be exactly reproduced.
This can be accomplished by solving a system with $2N$ equations that
can be written in matrix form: 
\be \label{eq:A.6}
\left(
  \ba{c}  Q_0 \\ Q_1 \\ ... \\ Q_{N-1} \\ Q_N \\ ... \\ Q_{2N-1}
  \ea \right)
=
\left( \ba{cccc} r_1^0 & r_2^0 & ... & r_N^0 \\
                  r_1^1 & r_2^1 & ... & r_N^1 \\
                   ...  &  ...  & ... &  ...  \\
                  r_1^{N-1} & r_2^{N-1} & ... & r_N^{N-1} \\
                  r_1^N & r_2^N & ... & r_N^N \\
                   ...  &  ...  & ... &  ...  \\
                  r_1^{2N-1} & r_2^{2N-1} & ... & r_N^{2N-1}
       \ea \right) \;\;
\left(
  \ba{c}  c_1 \\ c_2 \\ ... \\ c_N
  \ea \right)
\ee
It should be noted that in this system, both the $c_i$ and the $r_i$
are unknown. To solve it, the system must be partitioned in two, one
corresponding to the upper half equalities and the other to the lower
half ones. By defining:
\be \label{eq:A.7}
\ba{lcr}
\mathbf{Q} \equiv
\left(
  \ba{c}  Q_0 \\ Q_1 \\ ... \\ Q_{N-1}
  \ea \right)  \;\;\;\;\;\;\;\;\;\;\;\;
   &
\mathbf{Q'} \equiv
\left(
  \ba{c} Q_N \\ Q_{N+1} \\ ... \\ Q_{2N-1}
  \ea \right)
   &
  \;\;\;\;\;\;\;\;\;\;\;\;
\mathbf{c} \equiv
\left(
  \ba{c}  c_1 \\ c_2 \\ ... \\ c_N
  \ea \right)
\ea
\ee

\be \label{eq:A.8}
\ba{cc}
\mathbf{T} \equiv
\left( \ba{cccc}   1   &   1   & ... &   1   \\
                  r_1   & r_2   & ... & r_N   \\
                   ...  &  ...  & ... &  ...  \\
                  r_1^{N-1} & r_2^{N-1} & ... & r_N^{N-1}
       \ea \right)
   \;\;\;\;\;\;\;\;\;\;\;\;
&
\mathbf{T'} \equiv
\left( \ba{cccc}  r_1^N & r_2^N & ... & r_N^N   \\
                   r_1^{N+1} & r_2^{N+1} & ... & r_N^{N+1}   \\
                   ...  &  ...  & ... &  ...  \\
                   r_1^{N-1} & r_2^{N-1} & ... & r_N^{N-1}
       \ea \right)
\ea
\ee
it is clear that Eq.\ (\ref{eq:A.6}) is equivalent to:
\be \label{eq:A.9}
\ba{lr}
\mathbf{Q} = \mathbf{T} \; \mathbf{c}
   \;\;\;\;\;\;\;\;\;\;\;\;
&
\mathbf{Q'} = \mathbf{T'} \; \mathbf{c}
\ea
\ee
Solving for the first one:
\be \label{eq:A.10a}
\mathbf{c} = \mathbf{T}^{-1} \; \mathbf{Q}
\ee
and replacing in the second one, it follows:
\be \label{eq:A.10b}
\mathbf{Q'} = \mathbf{T'} \; \mathbf{T}^{-1} \; \mathbf{Q}
\ee
To attain directly the general solution of this latter equation can be
rather difficult. However, one can start by the first cases ($N = 1, 2,
...$) and proceed by induction. In this way, it can be found that the
set $\{ r_i \}_{i=1}^N$ that fulfills the equation conincides with the
roots of the $N$-th degree polynomial:
\be \label{eq:A.11}
\sum_{i=0}^N a_i^{(n)} \; r^i = 0 \;\;\;\;\;\;\;\;\;\;\;\; a_N^{(N)} = 1
\ee
where the coefficients, $a_i^{(N)}$ are the solutions of the linear
system:
\be \label{eq:A.12}
\left(
  \ba{c}  Q_N \\ Q_{N+1} \\ ... \\ Q_{2N-1}
  \ea \right)
= -
\left( \ba{cccc} Q_0 & Q_1 & ... & Q_{N-1} \\
                  Q_1 & Q_2 & ... & Q_N \\
                   ...  &  ...  & ... &  ...  \\
                  Q_{N-1} & Q_N & ... & Q_{2N-2} \\
       \ea \right) \;\;
\left(
  \ba{c}  a_0^{(N)} \\ a_1^{(N)} \\ ... \\ a_{N-1}^{(N)}
  \ea \right)
\ee
Finally, once the $r_i$ are obtained by solving Eq.\ (\ref{eq:A.11}), the
$c_i$ can be readily attained by Eq.\ (\ref{eq:A.10a}) and the values of
$\alpha_i$ follow from (\ref{eq:A.4b}).

Gaussian expansions of the sigma part of the $B$ functions ranging from
a single Gaussian to 20 Gaussians have been obtained. 
The standard limits
for the three- and four-center integrals are $N-L < 9$.

\vspace{1cm}
{\bf Acknowledgment}

The authors gratefully acknowledge the financial support
from the Spanish Direcci\'on General de Investigaci\'on
Cient\'{\i}fica y T\'ecnica and the German Deutscher
Akademischer Austauschdienst, num. of projects (HA1995-0034)
and (HA1996-0013). H.H.H.H. thankfully acknowledges support
by the Fonds der Chemischen Industrie.

\vspace{1cm}
{\bf Dedication}

We dedicate this article to Professor Klaus Ruedenberg with thankful appreciation of
his pioneering work on molecular integrals with exponential-type basis 
functions in molecular calculations.

\renewcommand{\baselinestretch}{1.5}

\begin{table}[htp]
\begin{center}
\caption{Least-squares errors$^a$, $\Delta^2$, in the Gaussian
expansions of $\hat{k}_{n-1/2}(r)$ obtained with the method of moments.}

\begin{tabular}{|c|cccccccc|}
\hline \hline
& & & & & & & & \\
\multicolumn{1}{|c|}{Number of} & \multicolumn{8}{c|}{$n$ index of the
$\hat{k}_{n-1/2}$ function} \\
\multicolumn{1}{|c|}{Gaussians} & & & & & & & & \\
& 1 & 2 & 3 & 4 & 5 & 6 & 7 & 8\\
& & & & & & & & \\
\hline
& & & & & & & & \\
1 & 2.7(-1)&5.9(-2) & 2.5(-2) & 1.4(-2) & 8.9(-3) & 6.1(-3) & 4.5(-3) & 3.4(-3) \\
2 & 3.2(-2)&3.6(-3) & 8.7(-4) & 3.1(-4) & 1.3(-4) & 7.0(-5) & 3.9(-5) & 2.4(-5) \\
3 & 7.1(-3)&4.6(-4) & 7.2(-5) & 1.8(-5) & 5.6(-6) & 2.2(-6) & 9.5(-7) & 4.6(-7) \\
4 & 2.0(-3)&8.4(-5) & 8.8(-6) & 1.5(-6) & 3.7(-7) & 1.1(-7) & 3.9(-8) & 1.5(-8) \\
5 & 7.0(-4)&1.9(-5) & 1.4(-6) & 1.8(-7) & 3.4(-8) & 8.0(-9) & 2.3(-9) & 7.4(-10)\\
6 & 2.8(-4)&5.0(-6) & 2.7(-7) & 2.7(-8) & 3.9(-9) & 7.3(-10)& 1.7(-10)& 4.7(-11)\\
7 & 1.2(-4)&1.5(-6) & 6.0(-8) & 4.6(-9) & 5.3(-10)& 8.2(-11)& 1.6(-11)& 3.7(-12)\\
8 & 5.7(-5)&4.9(-7) & 1.5(-8) & 9.0(-10)& 8.4(-11)& 1.1(-11)& 1.8(-12)& 3.5(-13)\\
9 & 2.8(-5)&1.8(-7) & 4.1(-9) & 2.0(-10)& 1.5(-11)& 1.6(-12)& 2.2(-13)& 3.8(-14)\\
10& 1.5(-5)&6.8(-8) & 1.2(-9) & 4.8(-11)& 3.0(-12)& 2.7(-13)& 3.2(-14)& 4.6(-15)\\
11& 8.3(-6)&2.8(-8) & 4.0(-10)& 1.2(-11)& 6.5(-13)& 5.0(-14)& 5.0(-15)& 6.4(-16)\\
12& 4.8(-6)&1.2(-8) & 1.4(-10)& 3.5(-12)& 1.5(-13)& 9.9(-15)& 8.7(-16)& 9.6(-17)\\
13& 2.9(-6)&5.4(-9) & 4.9(-11)& 1.0(-12)& 3.8(-14)& 2.1(-15)& 1.6(-16)& 1.6(-17)\\
14& 1.8(-6)&2.6(-9) & 1.9(-11)& 3.3(-13)& 1.0(-14)& 4.9(-16)& 3.3(-17)& 2.8(-18)\\
15& 1.1(-6)&1.3(-9) & 7.4(-12)& 1.1(-13)& 2.9(-15)& 1.2(-16)& 7.0(-18)& 5.3(-19)\\
16& 7.3(-7)&6.3(-10)& 3.0(-12)& 3.7(-14)& 8.6(-16)& 3.1(-17)& 1.6(-18)& 1.1(-19)\\
17& 4.9(-7)&3.3(-10)& 1.3(-12)& 1.4(-14)& 2.7(-16)& 8.5(-18)& 3.8(-19)& 2.3(-20)\\
18& 3.3(-7)&1.8(-10)& 5.8(-13)& 5.1(-15)& 8.7(-17)& 2.4(-18)& 9.6(-20)& 5.1(-21)\\
19& 2.3(-7)&9.8(-11)& 2.6(-13)& 2.0(-15)& 2.9(-17)& 7.1(-19)& 2.5(-20)& 1.2(-21)\\
20& 1.6(-7)&5.5(-11)& 1.2(-13)& 7.9(-16)& 1.0(-17)& 2.2(-19)& 6.9(-21)& 3.0(-22)\\
& & & & & & & & \\
\hline \hline
\end{tabular}
\end{center}
\end{table}

\begin{table}[htp]
\begin{center}
\caption{Single-Zeta BTO exponents.}

\begin{tabular}{|c|r|r|r|r|r|r|r|r|}
\hline 
Orbital  & Zn & S & B & C & N & O & F & H \\ 
\hline
1s &  28.979194 & 15.396775 & 4.649767 & 5.636105  & 6.621925  & 7.607778 & 8.593356 & 1.000000 \\
2s &   9.212368 &  4.468108 & 1.076139 & 1.346562  & 1.612481  & 1.885508 & 2.154463 &          \\
3s &   4.615722 &  1.723750 & 1.226030 & 1.581274  & 1.929475  & 2.238550 & 2.561510 &          \\
4s &   0.966290 &  5.987867 &          &           &           &          &          &          \\
2p &  13.015418 &  1.684294 &          &           &           &          &          &          \\
3p &   4.754359 &           &          &           &           &          &          &          \\
3d &   4.660219 &           &          &           &           &          &          &          \\
\hline 
\end{tabular}
\end{center}
\end{table}

\begin{table}[htp]
\begin{center}
\caption{Double-Zeta BTO exponents.}

\begin{tabular}{|c|r|r|r|r|r|r|r|r|}
\hline 
Orbital  & Zn & S & B & C & N & O & F & H \\ 
\hline
1s & 41.443334 & 16.517740 & 7.809120 & 9.121695  & 7.308276  & 8.306577 & 9.356483&  1.200000\\
1s & 29.336793 & 11.035952 & 3.907381 & 5.156559  & 4.211759  & 5.330563 & 6.202314&  1.000000\\
2s & 23.592780 &  5.492259 & 4.173803 & 5.447762  & 3.434959  & 2.910067 & 3.089333&          \\
2s & 12.264347 &  4.467502 & 1.204488 & 1.495602  & 1.724930  & 1.873454 & 2.040237&          \\
3s & 11.550144 &  2.255074 & 2.213227 & 2.726540  & 3.240492  & 3.686133 & 4.174972&          \\
3s &  5.617435 &  1.566569 & 1.004047 & 1.255198  & 1.496401  & 1.655654 & 1.847191&          \\
4s &  4.425408 &  9.438960 &          &           &           &          &         &          \\
4s &  1.140164 &  5.089873 &          &           &           &          &         &          \\
2p & 19.000918 &  2.008415 &          &           &           &          &         &          \\
2p & 11.734571 &  1.206954 &          &           &           &          &         &          \\
3p &  5.624098 &           &          &           &           &          &         &          \\
3p &  3.506195 &           &          &           &           &          &         &          \\
3d &  7.339257 &           &          &           &           &          &         &          \\
3d &  3.134250 &           &          &           &           &          &         &          \\
\hline 
\end{tabular}
\end{center}
\end{table}

\begin{table}[htp]
\begin{center}
\caption{Geometries used for molecular calculations}

\begin{tabular}{|l|c|ll|} \hline \hline
\multicolumn{1}{|l|}{ } &
\multicolumn{1}{l|}{ } & 
\multicolumn{2}{c|}{ } 
\\
\multicolumn{1}{|c|}{Molecule} &
\multicolumn{1}{c|}{Geometry} &
\multicolumn{2}{c|}{Bond distances and angles$^a$} 
\\    \hline  
\multicolumn{1}{|c|}{} &
\multicolumn{1}{c|}{} &
\multicolumn{2}{c|}{} 
\\  BH$_3$      & Planar & R$_{BH}$ = 2.25  &  
\\  B$_2$H$_6$  & See fig.\ 1 & R$_{BH}$ = 2.26013 & R$^\prime_{BH}$ = 2.53037   
\\              &            & R$_{BB}$ = 3.35430 & $\angle_{HBB} = 119^\circ$ 
\\  CH$_4$      & Regular tetrahedron & R$_{CH}$ = 2.0665  &  
\\  C$_2$H$_2$  & Linear & R$_{HC}$ = 2.002 & R$_{CC}$ = 2.281  
\\  C$_2$H$_4$  & Planar & R$_{HC}$ = 2.02203 & R$_{CC}$ = 2.55116 
\\              &        & $\angle_{HCH} = 120^\circ$  &
\\  C$_2$H$_6$  & Alternate & R$_{HC}$ = 2.08250 & R$_{CC}$ = 2.91588 
\\        &   & $\angle_{HCH} = 109.32^\circ$ & $\angle_{HCC} = 109.62^\circ$
\\  HCN         & Linear & R$_{HC}$ = 2.0  & R$_{CN}$ = 2.187  
\\  SF$_6$      & Regular octahedron & R$_{SF}$ =  2.88769  &  
\\  Zn$_3$      & Equilateral triangle & R$_{ZnZn}$ = 5.03593  &  
\\
\multicolumn{1}{|l|}{ } &
\multicolumn{1}{l|}{ } & 
\multicolumn{2}{c|}{ }
\\
\hline \hline
\end{tabular}
\end{center}

\begin{center}
$^a$ Distances in a.u.\ and angles in degrees.
\end{center}
\end{table}

\begin{table}[htp]
\begin{center}
\caption{Electronic Energy RHF with Single-Zeta basis set of BTO.$^a$}

\begin{tabular}{|c|r|lr|lr|lr|}
\hline \hline
Molecule & time(direct.alg.) & 10G & time & 15G & time & 20G & time  \\ \hline
BH$_3$     & 0.04 & -26.2978760 & 1.6   & -26.2978758    & 6.1   & -26.2978762597   & 16.4    \\
B$_2$H$_6$ & 0.6  & -52.591815  & 41.3  & -52.591817     & 156.9 & -52.5918199842   & 410.4 \\
H$_2$O     & 0.1  & -75.6136703 & 0.5   & -75.61366972   & 1.9   & -75.6136697787   & 5.0     \\
HCN        & 0.2  & -92.4898386 & 2.6   & -92.48983974   & 9.7   & -92.4898398408   &
25.0    \\
CH$_4$     & 0.04 & -40.0629604 & 3.5   & -40.0629605    & 13.7  & -40.0629611447   &
36.3    \\
C$_2$H$_2$ & 0.3  & -76.5372092 & 6.3   & -76.5372103    & 24.4  & -76.5372106538   &
61.6    \\
C$_2$H$_4$ & 0.4  & -77.7418899 & 19.8  & -77.7418903    & 74.8  & -77.7418909888   &
192.4    \\
C$_2$H$_6$ & 0.6  & -78.970476  & 40.4  & -78.9704946    & 155.8 & -78.9704966047   &
404.1    \\
SF$_6$     & 9.9  & -988.79773  & 598.7 & -988.79379     & 2277.8& -988.7938448335  &
5836.4 \\
Zn$_3$     & 82.4 & -5306.54329 & 275.4 & -5306.54320060 & 614.4 & -5306.5432005659 &
1679.2 \\ \hline \hline
\end{tabular}
\end{center}

\begin{center}
$^a$ Energy in Hartrees.
\end{center}
\end{table}

\begin{table}[htp]
\begin{center}
\caption{Electronic Energy RHF with Double-Zeta basis set of BTO.$^a$}

\begin{tabular}{|c|r|lr|lr|lr|}
\hline \hline
Molecule & time(direct.alg.) & 10G & time & 15G & time & 20G & time  \\
\hline
BH$_3$     & 0.5   & -26.373450 & 23.1   & -26.3734463  & 90.0   & -26.3734465526   &
234.7 \\
B$_2$H$_6$ & 5.3   & -52.757514 & 595.0  & -52.757528   & 2290.9 & -52.7575305959   &
5967 .9   \\
H$_2$O     & 0.3   & -75.998708 & 7.3    & -75.99870643 & 28.0   & -75.9987064604   &
73.0 \\
HCN        & 2.5   & -92.801141 & 36.4   & -92.8011484  & 135.6  &
-92.8011486868 & 350.0 \\
CH$_4$     & 0.9   & -40.182804 & 51.6   & -40.1828002  & 202.8  &
-40.1828008395 & 532.5 \\
C$_2$H$_2$ & 3.0   & -76.773130 & 89.1   & -76.7731380  & 336.4  & -76.7731382182
& 872.2 \\
C$_2$H$_4$ & 4.2   & -77.99422  & 281.3  & -77.9942558  & 1070.7 &
-77.9942574766 & 2791.2 \\
C$_2$H$_6$ & 5.1   & -79.197399 & 589.0  & -79.1974595  & 2255.7 &
-79.1974630394 & 5853.2 \\
SF$_6$     & 141.5 & -993.7002  & 8683.5 & -993.70374   & 33498.9&
-993.7038527430 & 87495.1 \\
Zn$_3$     & 744.6 & -5332.8974 & 2007.0 & -5332.89797  & 6802.8 &
-5332.8979948489 & 16198.3 \\ \hline \hline
\end{tabular}
\end{center}

\begin{center}
$^a$ Energy in Hartrees.
\end{center}
\end{table}

\begin{table}[htp]
\begin{center}
\caption{RHF electronic energy with both Slater and $B$ basis sets.$^{a}$}

\begin{tabular}{|l|rrrr|}
\hline \hline
   & Bessel(SZ) \cite{bases} & Slater(SZ) \cite{Clementi} &
Bessel(DZ) \cite{bases} & Slater(DZ) \cite{koga} \\ \hline
$BH_3 $ &-26.297876    & -26.319008   & -26.373450    & -26.375009  \\
$B_2H_6$ &-52.591815    & -52.630434   & -52.757514    & -52.758232  \\
$H_2O $ &-75.613670    & -75.687535   & -75.998708    & -76.000535  \\
$HCN $ &-92.489839    & -92.572795   & -92.801148    & -92.833977  \\
$CH_4 $ &-40.062960    & -40.101608   & -40.182800    & -40.183596  \\
$C_2H_2$ &-76.537209    & -76.596599   & -76.773130    & -76.803955  \\
$C_2H_4$ &-77.741890    & -77.810378   & -77.99422     & -78.004891  \\
$C_2H_6$ &-78.970476    & -79.041644   & -79.197399    & -79.199828  \\
$SF_6 $ &-988.79773    & -989.887733  & -993.7002     & -993.705704 \\
$Zn_3 $ &-5306.54329   & -5313.330879 & -5332.8974    & -5332.907860\\
\hline \hline
\end{tabular}
\end{center}

\begin{center}
$^a$ Energy in Hartrees.
\end{center}
\end{table}

\newpage

\pagestyle{empty}                                   
\begin{figure}
\setlength{\unitlength}{1.cm}                       
\begin{picture}(10,10)(0,0)

\put(3,3){\ \psfig{file=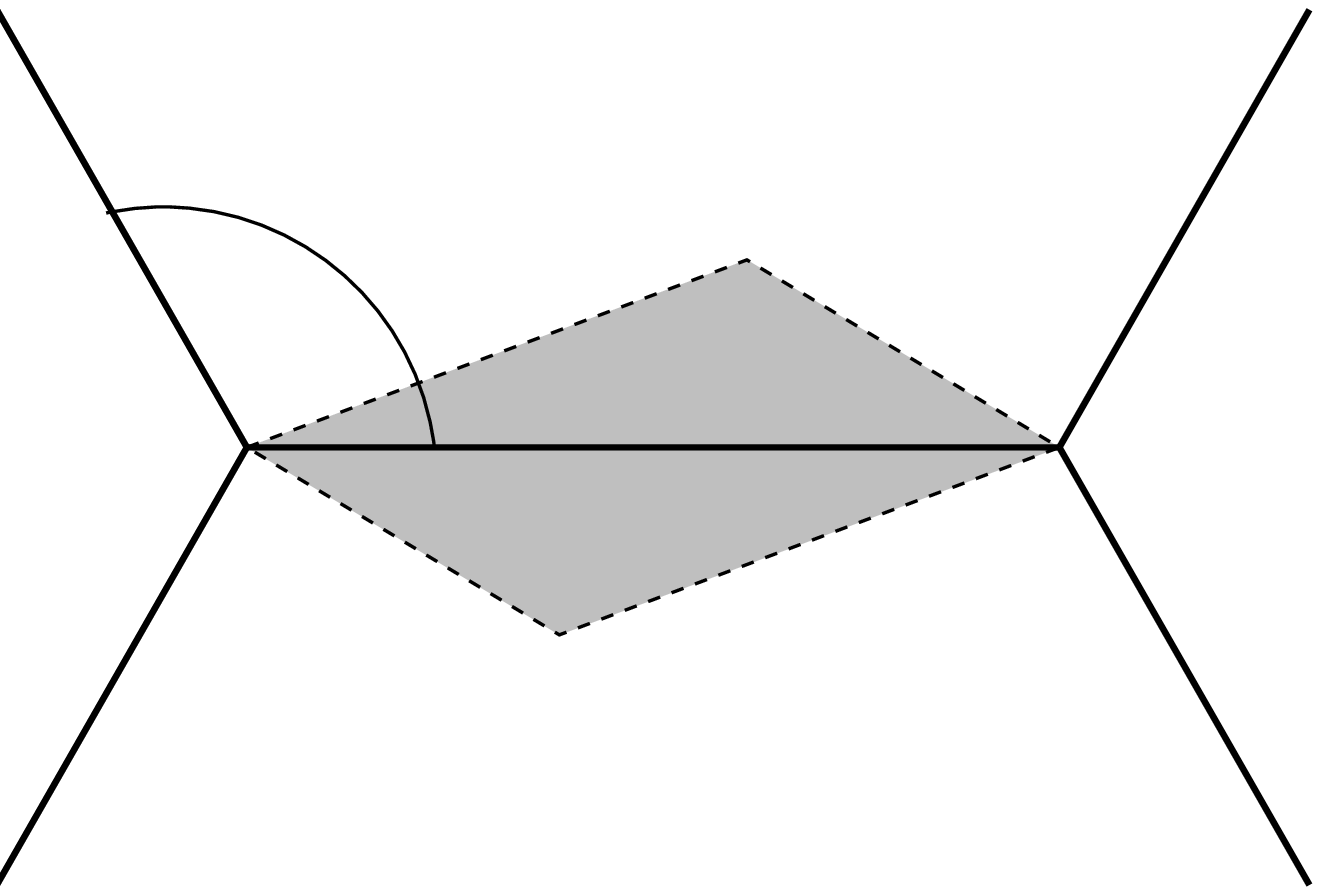,height=6.cm,width=9.cm,clip=}}

\put(2.4,9.){\large{\bf{H}}}
\put(2.4,2.9){\large{\bf{H}}}
\put(12.5,9.){\large{\bf{H}}}
\put(12.5,2.9){\large{\bf{H}}}
\put(4,5.8){\large{\bf{B}}}
\put(11,5.8){\large{\bf{B}}}
\put(6.8,4.0){\large{\bf{H}}}
\put(7.9,7.6){\large{\bf{H}}}
\put(5.5,7.6){\large{$\angle_{HBB}$}}
\put(11.5,7.5){\large{$R_{HB}$}}
\put(8.7,4.7){\large{$R^\prime_{HB}$}}
\put(7.3,6.3){\large{$R_{BB}$}}

\end{picture}                                       
\end{figure}

\end{document}